\documentclass[usenatbib,usegraphicx]{mn2e}

\usepackage{amsmath}
\usepackage{subfigure}

\newcommand{\Rj}{\hbox{$R_{\mathrm{J}}$\,}}

\newcommand{\Rs}{\hbox{$R_{*}$\,}}

\newcommand{\Rp}{\hbox{$R_{\mathrm{p}}$\,}}

\newcommand{\rmsub}[2]{#1_{\rm #2}}

\title[Correlated noise in SuperWASP data]{The impact of correlated noise on SuperWASP
detection rates for transiting extra-solar planets}

\author[A.M.S. Smith et al.]{A.M.S. Smith$^{1}$\thanks{E-mail:
amss@st-and.ac.uk},
A. Collier Cameron$^{1}$,
D.J. Christian$^{2}$,
W.I. Clarkson$^{3,4}$,
B. Enoch$^{3}$,
\newauthor
A. Evans$^{5}$,
C.A. Haswell$^{3}$,
C. Hellier$^{5}$,
K. Horne$^{1}$,
J. Irwin$^{6}$,
S.R.~Kane$^{1,7}$,
\newauthor
T.A. Lister$^{1,5}$,
A.J. Norton$^{3}$,
N. Parley$^{3}$,
D.L. Pollacco$^{2}$,
R. Ryans$^{2}$,
I. Skillen$^{8}$,
\newauthor
R.A. Street$^{2}$,
A.H.M.J. Triaud$^{1}$,
R.G.~West$^{9}$,
P.J. Wheatley$^{10}$,
and D. M. Wilson$^{5}$ \\
$^{1}$SUPA\thanks{Scottish Universities Physics Alliance}, School of Physics \& Astronomy, University of St. Andrews, North Haugh, 
St. Andrews, Fife, KY16 9SS, UK,  \\
$^{2}$Astrophysics Research Centre, Main Physics Building, School of Mathematics \&
Physics, Queen's University, Belfast, BT7 1NN, UK,\\ 
$^{3}$Department of Physics \& Astronomy, The Open University, Milton Keynes, 
MK7 6AA, UK,  \\
$^{4}$Space Telescope Science Institute (STScI), 3700 San Martin Drive,
Baltimore, MD~21218, USA,\\
$^{5}$Astrophysics Group, School of Chemistry \& Physics, Keele University, 
Staffordshire, ST5 5BG, UK,  \\
$^{6}$Institute of Astronomy, University of Cambridge, Madingley Road, 
Cambridge, CB3 0HA, UK, \\ 
$^{7}$University of Florida, PO~Box~112005, 211 Bryant Space Science Center,
Gainesville, FL, USA.\\
$^{8}$Isaac Newton Group of Telescopes, Apartado de correos 321,
E-38700 Santa Cruz de la Palma, Tenerife, Spain,\\ 
$^{9}$Department of Physics \& Astronomy, University of Leicester, Leicester, 
LE1 7RH, UK,\\
$^{10}$Department of Physics, University of Warwick, Coventry CV4 7AL, UK.\\
}

\begin{document}
\maketitle

\begin{abstract}

We present a model of the stellar populations in the fields observed by one of the SuperWASP-N cameras in the 2004
observing season. We use the Besan\c con Galactic model to define the range of stellar
types and metallicities present, and populate these objects with
transiting extra-solar planets using the metallicity relation of \cite{F&V05}.

We investigate the ability of SuperWASP to detect these planets in the presence of
realistic levels of correlated systematic noise (`red noise'). We find that the number of planets that transit with a signal-to-noise ratio of 10 or more
increases linearly with the number of nights of observations. Based on a simulation of
detection rates across 20 fields observed by one camera, we predict that a total of $18.6
\pm 8.0$ planets should be detectable from the SuperWASP-N 2004 data alone. The best way to
limit the impact of co-variant noise and increase the number of detectable planets is to
boost the signal-to-noise ratio, by 
increasing the number of observed transits for each candidate transiting planet. This
requires the observing baseline to be increased, by spending a second
observing season monitoring the same fields.

\end{abstract}
\begin{keywords}
techniques: photometric -- planetary systems

\end{keywords}
\section{Introduction}

Since the discovery of 51 Peg b by \cite{M&Q95}, a total of 181 extra-solar
planets around 155 stars (as at 2006 May) have been discovered\footnote{See the {\it Extrasolar
Planets Encyclopaedia} at http://exoplanet.eu/catalog.php}, almost exclusively
by radial velocity measurements. It was only with the
discovery that HD209458b transits its host star \citep{Charbonneau-etal00} that
both the mass and radius of an extra-solar planet were first determined. A
further eight transiting extra-solar planets have been discovered in the last
few years. Most of these planets have been discovered with the OGLE transit
survey (e.g. \citealt{Konacki-etal03}), which is a deep-field survey. SuperWASP
(Wide Angle Search for Planets) is a wide-field transit survey, which is
capable of detecting planetary transits of stars that are bright enough for high
precision radial velocity follow-up observations.

Transit surveys have the potential to find numerous hot Jupiter-like planets
(HJs), that is Jupiter-sized planets which orbit close to their host
star with periods of just a few days. The probability that a HJ system is
aligned such that transits will occur is about 0.1; which is much more
favourable than planets orbiting at greater distances (for an Earth-like orbit,
the probability of transit alignment is 0.0046).

Crucially, the detection of transiting planets allows the determination of the
average density of the planet. This is of interest for planet formation models,
particularly since HD209458b appears to have an anomalously low density
\citep{Charbonneau-etal00}, which is not well understood.

Previous attempts have been made to estimate the expected detection rates of transiting
HJs by shallow, wide-field transit surveys similar to SuperWASP (e.g. \citealt{Brown03}). Based on an observing pattern
consisting of 38 nights of observations
spread over 91 days and a requirement that three or more transits are observed, Brown
calculated the detection
rate of HJs producing a transit of depth 1 per cent or greater to be 0.39 per $10^4$
stars. Brown also estimates that, for the same observing window function, 4.51 false
alarms will be detected per $10^4$ stars -- indicating that only eight per cent of
transit signals detected will be produced by planets. These planetary transit
`impostors' are grazing eclipsing binaries and eclipsing binaries diluted by light from
a third star (either a field star or the third member of a triple system).

The SuperWASP survey has the potential to define the population of extra-solar planets which
transit nearby bright stars ($V < 13$). In order for the results of SuperWASP to be properly
interpreted, it is essential that the selection effects that operate in the survey are well
understood. In this work, we use the findings of \cite{Pont06} and Pont, Zucker \& Queloz
(2006) to estimate SuperWASP's
detection rate in the presence of realistic levels of systematic red noise. We find that in
order to detect a significant number of transiting planets, the
existence of red noise necessitates much longer observing baselines than previously thought.

\section{Observations}

SuperWASP is a wide-field transit survey with instruments at the Isaac Newton Group
on La Palma (SuperWASP-N) and at the South African Astronomical Observatory, near
Sutherland, RSA (SuperWASP-S). SuperWASP-N starting observing in 2004 and
SuperWASP-S was commissioned in early 2006.

Each SuperWASP installation comprises an array of eight 200mm f/1.8 Canon camera
lenses\footnote{SuperWASP-N
operated for the 2004 season with only five cameras}, each with a four mega-pixel CCD recording a 7.8\degr ~by 7.8\degr ~field.
These eight cameras are mounted on a fork mount, which is housed in an enclosure
with a hydraulically operated roof. The enclosure also contains the operating
computers and a weather station.

The observing strategy for the 2004 season was designed to maximise the number
of stars that could be hunted for transits -- overcrowded fields near the
galactic plane were avoided \citep{Christian-etal05}. Up to eight sets of fields were observed
at a time, with around a minute spent on each, including a 30 s exposure, 4
s of read-out time, and
slewing, giving measurements about 7 - 8 minutes apart.

\section{Data Reduction}

Since each SuperWASP exposure generates an 8.4 Mb FITS file, a large quantity of
data is stored and processed; this is done using a custom-written data
reduction pipeline \citep{Pollacco-etal06}. The pipeline uses the Tycho 2
\citep{Tycho} and USNO B
\citep{USNOB}
catalogues to prepare an astrometric solution for each field. Aperture
photometry is performed on all objects, using three different sized apertures
and a blending index is assigned for each object. Airmass and CCD position
trends are removed from the data before it is archived at the University of
Leicester.

We search for transits in the data using a hybrid Box Least-Squares and Newton-Raphson
algorithm, {\sc Hunter}, developed by \cite{Cameron-etal06}. Prior to searching for
transits with {\sc Hunter}, the 
{\sc SysRem} algorithm of \cite{Tamuz-etal05} is applied to the data to reduce
systematic errors. Four components of systematic error are removed with {\sc
SysRem}; it is thought that two of these are caused by airmass variations and
thermal changes in the camera focus, but the causes of the other two are unknown
\citep{Cameron-etal06}.

{\sc Hunter} searches a grid of transit
periods and epochs to fit a transit model to the data. The $\chi^2$ statistic is
used to determine whether a best-fitting model includes a transit and, if so, to
determine the best-fitting transit parameters for each candidate. 

\section{`Colours' of noise in photometric data}

\cite{Pont06} demonstrated that there is likely to be significant co-variance
structure in the noise in data from ground-based photometric surveys, such as
SuperWASP. Previous forecasts of
the planet `catch' from such instruments (e.g. \citealt{Horne01})  have assumed that such
noise is un-correlated or `white' in nature. Pont suggests that the reduced
signal-to-noise caused by correlated or `red' noise can account for an
observed shortfall in transiting planet detections.

Noise consisting of white, independent, random noise combined with red, co-variant,
systematic noise is termed `pink', and, unlike white noise, cannot be removed by averaging
the data. Pont showed that systematic noise, correlated on time scales
equivalent to a
typical hot Jupiter transit ($\approx 2.5$ hours) cannot be ignored and indeed tends to be
the dominant type of noise for bright stars. It therefore seems likely that the
noise in SuperWASP data will be pink.

\section{Characterisation of SuperWASP noise}
\label{sec:rednoise}

The simplest method of establishing the level of correlated noise present in the
data is to compute a running average of the data over the $n$ data points
contained in a transit-length time interval  \citep{PZQ06}.
The transit duration
chosen here is 2.5 hours, which is the transit duration corresponding to a planet
orbiting a solar analogue, with a period of 2.6 days -- typical of a hot Jupiter.
Since exposures are taken at roughly 7 minute intervals, there are
about $n=20$ points in each interval.

If the noise is purely random, the RMS scatter in the average of $n$ data points should be 
$\sigma_\mathrm{w}=\sigma \sqrt{n}$, where $\sigma$ is the standard RMS of the whole lightcurve. If,
however, there is a systematic component in the noise, the RMS scatter of the
average of $n$ points will be greater than this.

The RMS scatter, $\sigma$, is calculated for each of the 822 stars determined to
be non-variable in the field
centred at 15h17min RA, +23\degr 26\arcmin ~dec for which lightcurves have been produced by the SuperWASP data
reduction pipeline. The noisiest 25 per cent of the
data points in each lightcurve, corresponding to measurements made around full
moon and during Sahara dust events, are excluded from the analysis. Stars are
determined to be variable, and excluded from the analysis, if
$\sqrt{\sigma_\mathrm{s}^2}>0.005~\mathrm{mag}$, where $\sigma_\mathrm{s}^2$ is
the variance caused by intrinsic stellar variability, derived from the $\chi^2$
statistic for a constant-flux model.

The running average,
$\sigma_\mathrm{r}$, over 20 points is also calculated for each of these stars,
with the same exclusions of the noisiest data and intrinsically variable stars.

\begin{figure}
\subfigure{\label{fig:non-Tamuz}
\includegraphics[angle=270,width=8.25cm]{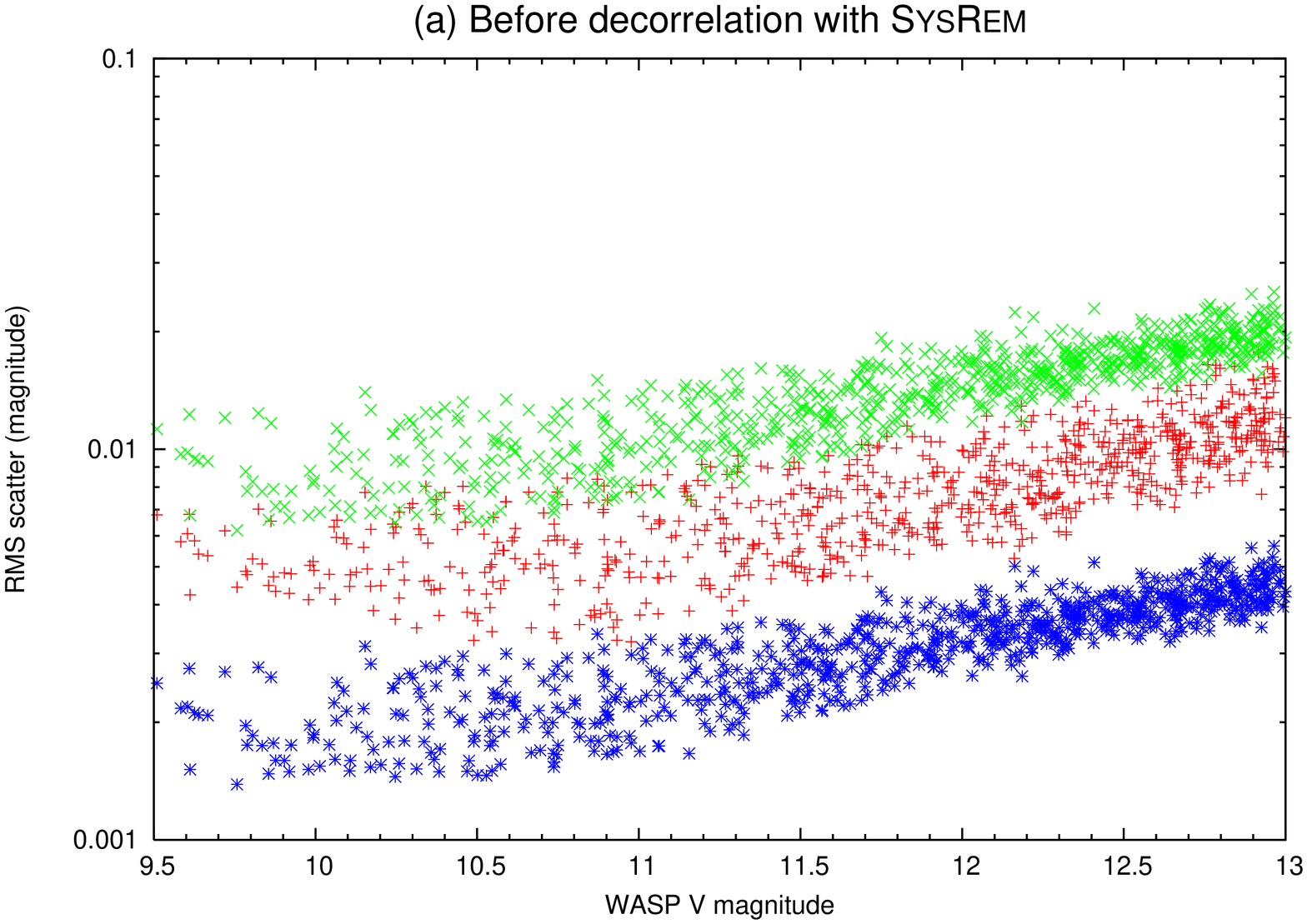}
}
\subfigure{\label{fig:Tamuz}
\includegraphics[angle=270,width=8.25cm]{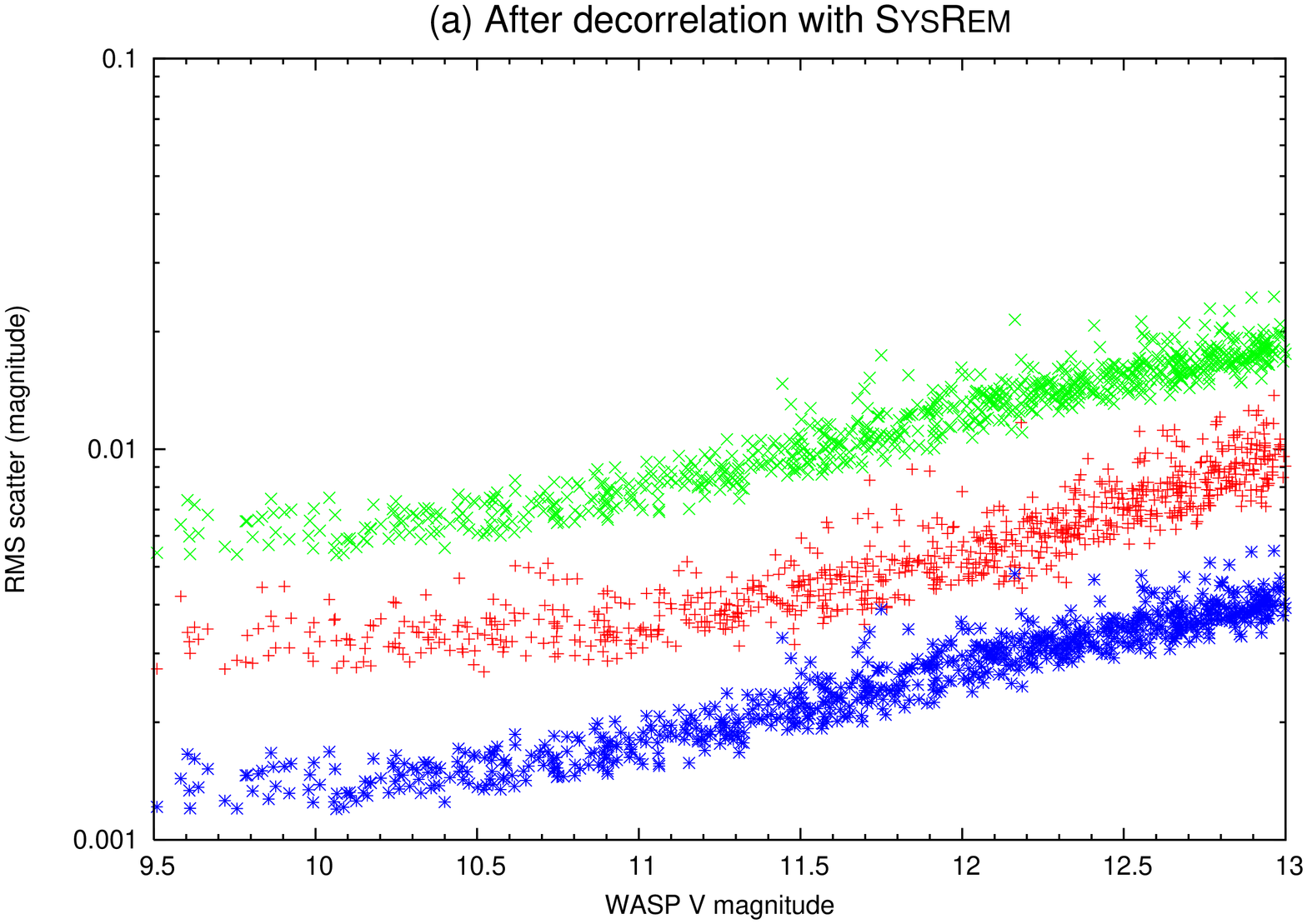}
}
\caption{RMS scatter versus magnitude for non-variable stars in the field centred
at 15h17min RA, +23\degr 26\arcmin ~dec both prior to (a), and after
(b) decorrelation with {\sc SysRem}. The upper curve shows the RMS scatter of the
lightcurve of each object, $\sigma$. The middle curve shows the scatter,
$\sigma_\mathrm{r}$, after a 
moving average over a 2.5 hour time interval (20 data points) was calculated. The
lower curve shows the RMS scatter divided by 20, $\sigma_\mathrm{w}$.
}
\protect\label{fig:RMS}
\end{figure}

Both $\sigma_\mathrm{w}$ and $\sigma_\mathrm{r}$ are calculated prior to, and
after, de-correlation with {\sc SysRem}. These quantities, and $\sigma$, are
plotted against magnitude for the 822 non-variable stars in the field in
Fig.\ref{fig:RMS}.

As indicated by the differences between Figs. \ref{fig:non-Tamuz} and
\ref{fig:Tamuz}, the {\sc SysRem} algorithm is highly effective at reducing the
levels of systematic noise present in the data. Fig. \ref{fig:Tamuz} also shows,
however, that not all correlation in the noise is removed by {\sc SysRem}. If that
were the case, the $\sigma_\mathrm{r}$ curve would lie over the
$\sigma_\mathrm{w}$ curve. Instead, the $\sigma_\mathrm{r}$ curve lies
higher than $\sigma_\mathrm{w}$, and flattens out at about 3~mmag for bright
($V$=9.5) stars, indicating that systematic trends of this magnitude are present in
the data on a 2.5 hour timescale.

\section{Simulated planet catch}

We model the objects in the 36 fields viewed by one of the SuperWASP cameras in the 2004
season, by using the Besan\c con model of the Galaxy
\citep{Robin-etal03} to generate a star catalogue of stars with $9.5<V<13.0$ for each of
the fields. Planets are then assigned to stars that are of spectral class F, G or
K and luminosity class IV or V on the basis of their metallicity, using the
planet-metallicity relation of \cite{F&V05}:
\begin{equation}
\mathcal{P}(\mathrm {planet}) =0.03\times10^{2.0[\mathrm{Fe/H}]}
\protect\label{eqn:F&V}
\end{equation}
where $\mathcal{P}(\mathrm {planet})$ is the probability that a particular star of
metallicity [Fe/H] is host to a planet. The above equation is used for stars which have metallicities in the range
-0.5$<$[Fe/H]$<$0.5. For stars with [Fe/H]$<-0.5$, $\mathcal{P}(\mathrm
{planet})=~0.003$, and
for [Fe/H]$>0.5$, $\mathcal{P}(\mathrm{planet})= 0.3$.

Only stars of spectral type F,G and K are allocated a non-zero planet hosting
probability, since the \cite{F&V05} equation is based upon radial
velocity observations of stars of this type only. This does not pose a significant
problem, however, as early-type stars are not numerous and have radii that are too
great for transit detection. M-type stars are not particularly numerous in the
Besan\c con-generated catalogues either (about 2.8 per cent of the stars are of
type M or later), because the sample is limited by apparent
magnitude, not volume. Although the transit signal produced by a Jupiter-like
planet orbiting an M dwarf star will be greater than that produced by, say, a
G dwarf star, M dwarfs are thought to be less likely to harbour giant planets
\citep{Adams05}. Similarly, only subgiants and dwarfs are considered planet hosts,
since any planet orbiting a giant star would produce an insufficiently deep
transit signature to be detected.

The probability that a star hosts a transiting planet is calculated using
\begin{equation}
\mathcal{P}(\mathrm {transit}) =\mathcal{P}(\mathrm {planet})\times
\mathcal{P}(\mathrm {alignment})
\end{equation}
where $\mathcal{P}(\mathrm {alignment})$ is the probability of a given planetary
system being aligned with respect to the line of sight such that a transit can be
observed. This is given by
\begin{equation}
\mathcal{P}(\mathrm {alignment})=\mathrm{arctan}\left(\frac{\Rs+\Rp}{a}\right)
\approx \mathrm{arctan}\left(\frac{\Rs}{a}\right)
\end{equation}
where $\Rs$ is the stellar radius, $\Rp$ the planetary radius and $a$ the
semi-major axis of the system. The semi-major axis of each potential planetary system in the model is drawn
randomly from a distribution that is uniform in log ($a$), between $0.02 < a <
5.25$ au. 

\begin{figure}
\includegraphics[angle=270,width=8.4cm]{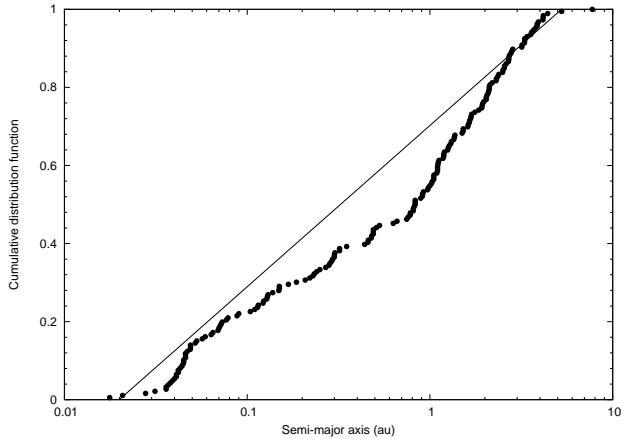}
\caption{Cumulative distribution function of extra-solar planetary semi-major axis. The
filled circles represent the 186 planets discovered by radial velocity means, as at 2006
September 15. The solid line is the distribution (uniform in log ($a$) for
 $0.02~<~a~<~5.25$~au) used in our model.}
\label{fig:cdf}
\end{figure}

This simple log-flat distribution of semi-major axes is compared to the actual
distribution of $a$ amongst extra-solar planets discovered by Doppler surveys in
Fig. \ref{fig:cdf}. Our distribution appears to be a poor fit to the observed distribution in
the regime favoured by transit surveys ($a \la 0.05$ au), which may lead to an
overestimation of the number of very short-period planets. The distributions,
however, closely agree on the fraction of planets with $a \le$ 0.05 au - 14 per cent
of the Doppler survey planets have $a \le$ 0.05 au, while the model distribution
gives 16 per cent.

On the basis of the probability $\mathcal{P}(\mathrm {transit})$, for each star, 
it is determined whether or not a star hosts a transiting planet. It is assumed,
for simplicity, that all planets have a radius, $\Rp$, equal to that of Jupiter,
$\Rj$. The depth of transit, $\Delta m$, is determined from the
equation of \cite{T&S},
\begin{equation}
\Delta m \approx 1.3 \left(\frac{\Rp}{\Rs}\right)^2
\label{eqn:depth}
\end{equation}
The factor of 1.3 in the above equation takes account of the effect of
stellar limb-darkening, and although this assumes a central transit, off-centre transits will only
have a slightly smaller limb-darkening factor \citep{T&S}.

A total of 355,429 stars are generated by the Besan\c con model in the 36 fields, 165586
(46.6 per cent) of which are of type F,G or K and class IV or V. The simulation described
above results in the allocation of a transiting planet to 329 of
these stars, although this number changes each time the simulation is run because it
relies on random numbers (see \S \ref{sec:no_planets} for discussion of this). The transit
depths of these 329 systems are calculated using equation \ref{eqn:depth}.

The detection thresholds for planet detection are determined by fitting lines to
the white and red noise curves of Fig. \ref{fig:Tamuz}. The data is modelled with a
constant term and a term which is inversely proportional to the flux of each object. This
leads to fitted lines of
the form $\sigma~=~c_1~+~c_2~\left(10^{0.4V}\right)$, where $c_1$ and $c_2$
are constants which are fitted for, and $V$ is the $V$-band magnitude. The values
of the constants $c_1$ and $c_2$, as determined by $\chi ^2$ minimisation, for red
and white noise are shown in Table \ref{tab:fitparas}.

\begin{table}
\centering
\caption{Fitted parameters for RMS scatter as a function of magnitude for the
white and red noise cases}
\label{tab:fitparas}
\begin{tabular}{lcc}
\hline
 & $c_1$ & $c_2$ \\
\hline
$\sigma _\mathrm{w}$ & $1.40 \times 10^{-3}$ & $1.96 \times 10^{-8}$ \\
$\sigma _\mathrm{r}$ & $2.88 \times 10^{-3}$ & $4.34 \times 10^{-8}$ \\
\hline
\end{tabular}

\medskip

All variables are defined in the text.
\end{table}

\begin{figure}
\includegraphics[angle=270,width=8.4cm]{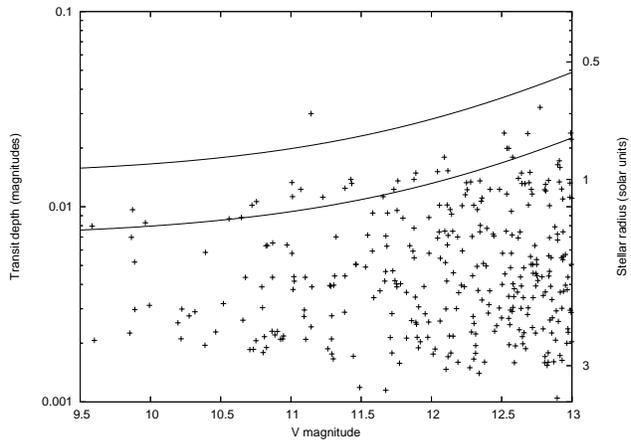}
\caption{Transit depth versus magnitude for 329 simulated transiting extra-solar 
planets (points), with 5-$\sigma$ detection thresholds for red (upper curve) and white
(lower curve) noise, for a single transit. If more transits are observed, the
thresholds are lowered by a factor of $\sqrt{n_\mathrm{trans}}$. Also indicated is the stellar radius, which is directly
related to the transit depth since all planets are of radius $\Rj$.}
\label{fig:depths}
\end{figure}

These functions, $\sigma _\mathrm{w}$ and $\sigma _\mathrm{r}$, of magnitude are the
1-$\sigma$ detection thresholds
for the white and red noise cases respectively. The transit depths of the 329
simulated transiting planets are plotted as a function of magnitude along with the
5-$\sigma$ detection thresholds for white and red noise in Fig. \ref
{fig:depths}. 28 of the 329 planets have transit depths greater than the
5-$\sigma$ detection threshold for white noise, but only one has a depth greater
than the equivalent threshold when red noise is considered.

\subsection{Signal-to-noise ratios}

Although Fig. \ref{fig:depths} provides a neat illustration of the reduction in planet
detection efficiency experienced when red noise is present in the data, an analysis of the
signal-to-noise ratio allows a fuller picture of the red noise problem to emerge.

The signal-to-noise ratio, $\rmsub{S}{red}$, is a function of the number of observations made
in-transit, as well as the depth of the transit. In the presence of red noise,
$\rmsub{S}{red}$ is 
given by
\begin{equation}
\rmsub{S}{red}=\frac{\Delta m\sqrt{n_\mathrm{trans}}}{\sigma_\mathrm{r}(V)}
\label{eqn:SNR}
\end{equation}
where $n_\mathrm{trans}$ is the number of transits observed.

$\rmsub{S}{red}$ is similar to the $S_r$ statistic of \cite{PZQ06}, which also considers
white noise. In the regime applicable to this work, where red noise dominates, $S_r$
simplifies to our expression for $\rmsub{S}{red}$.

$\rmsub{S}{red}$ is a criterion used in the selection of planetary transit
candidates for follow-up spectroscopic observations \citep{Cameron-etal06}; a
signal-to-noise ratio of 10 or greater is required in order for an object 
to be considered as a viable transit candidate. This is a compromise between allowing
too many false positive detections and rejecting large numbers of genuine transiting
planets.

The $\rmsub{S}{red}$ values of the simulated planets can be calculated using equation \ref{eqn:SNR}, since the
transit depth, $\Delta m$, and the RMS scatter as a function of magnitude for red noise,
$\sigma_\mathrm{r}(V)$, are already known. $n_\mathrm{trans}$ depends on the observing
pattern and observing baseline used; here it is calculated for each simulated transiting
planet using the observation times of three SuperWASP fields, the details of which are
summarised in Table \ref{tab:fields}.

\begin{table}
\centering
\caption{The number of nights of observations for the three fields used in the
signal-to-noise ratio analysis.}
\label{tab:fields}
\begin{tabular}{lc}
\hline
 SuperWASP field ID& No. of nights \\
\hline
SW1143+3126 & 51 \\
SW0044+2826 & 80 \\
SW1743+3126 & 130 \\
\hline
\end{tabular}
\end{table}

Each simulated planet is assigned a random epoch which is combined with the orbital period
to produce an ephemeris for each simulated planet.

\begin{figure}
\subfigure{\label{fig:51snr}
\includegraphics[angle=270, width=8.25 cm]{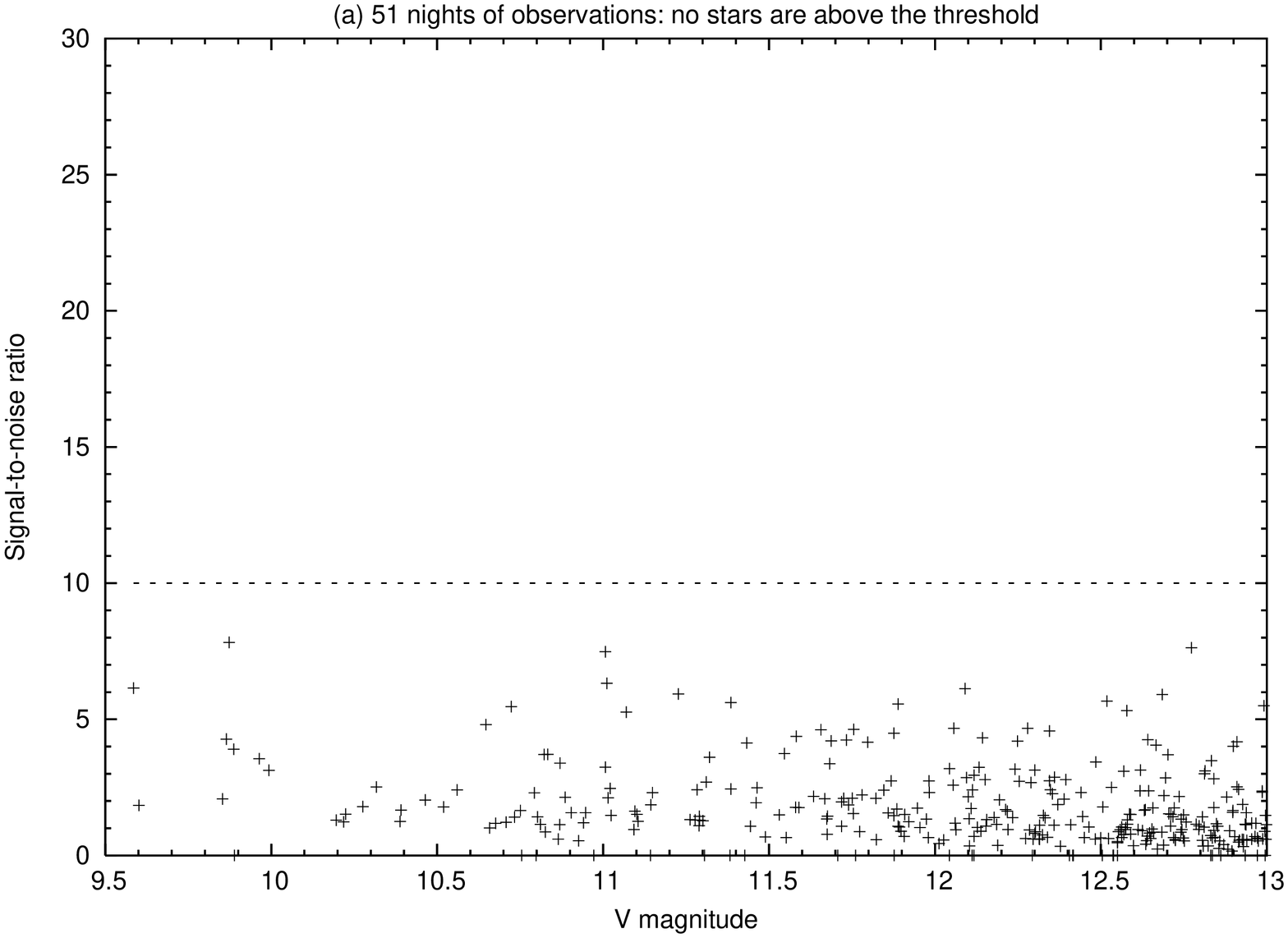}
}
\subfigure{\label{fig:80snr}
\includegraphics[angle=270, width=8.25 cm]{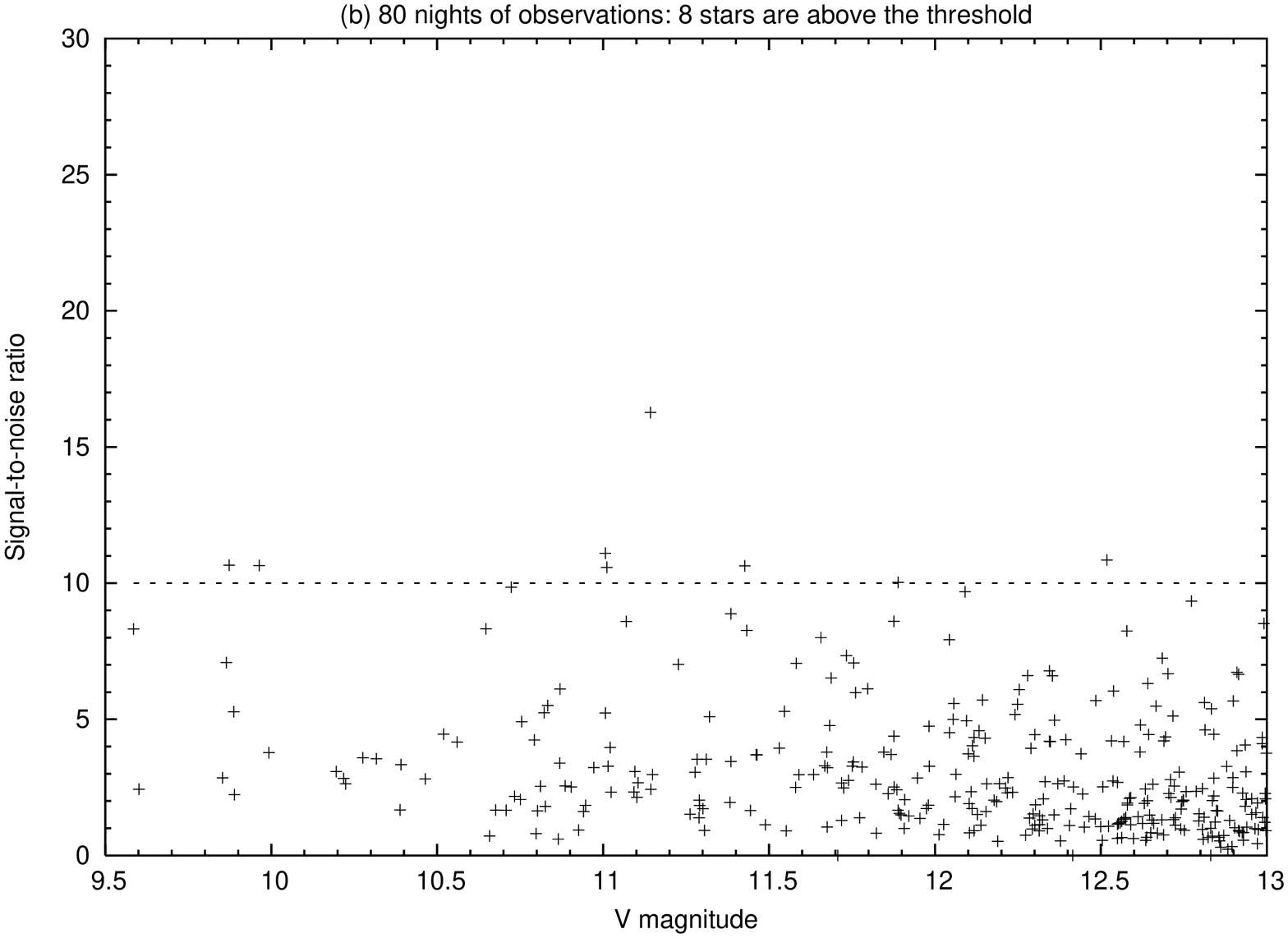}
}
\subfigure{\label{fig:130snr}
\includegraphics[angle=270, width=8.25 cm]{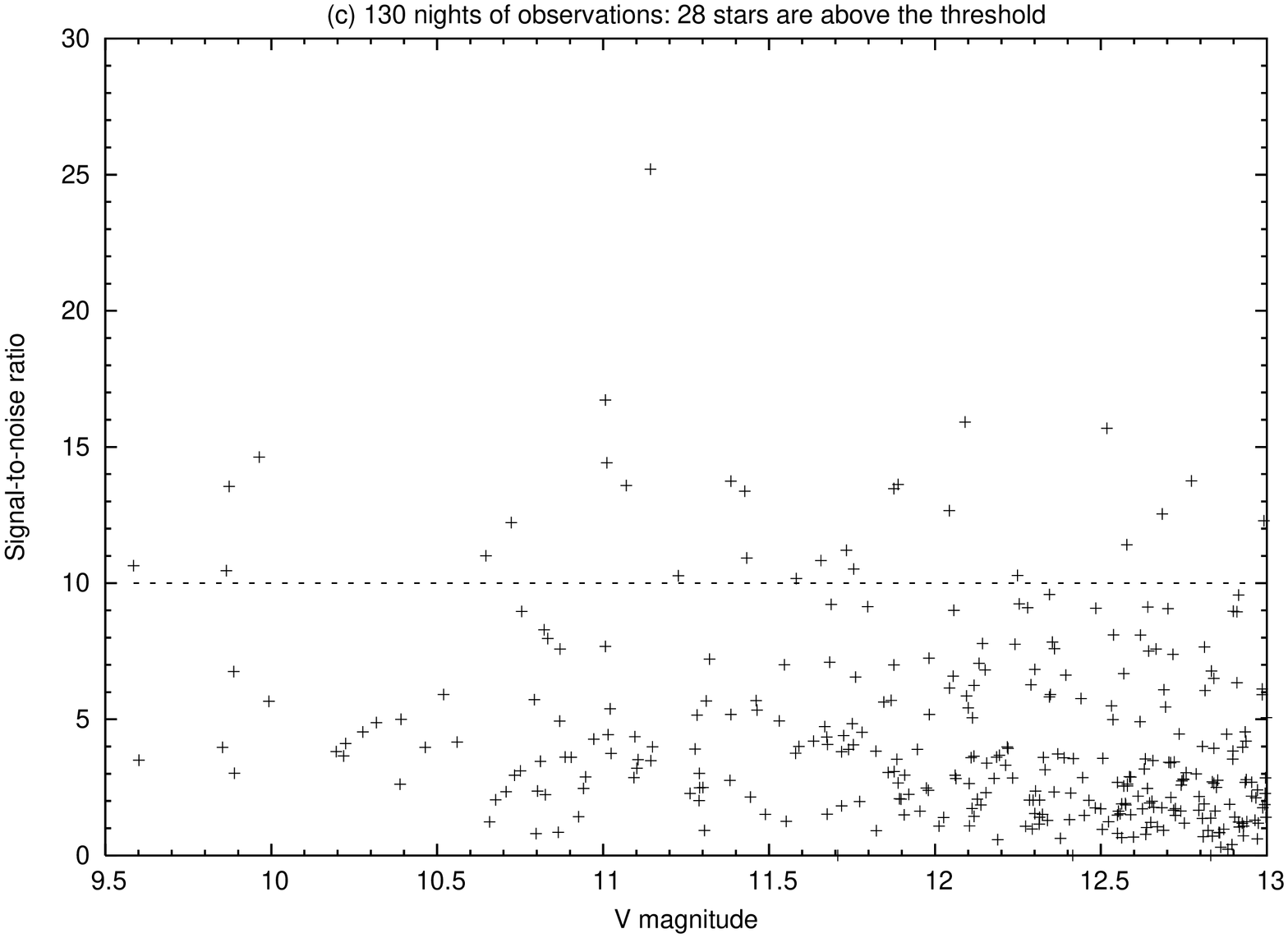}
}
\caption{Signal-to-noise ratio versus magnitude for 329 simulated transiting
extra-solar planets, for 51, 80 and 130 nights of data. The dotted line indicates a
signal-to-noise ratio of 10, the threshold used when compiling transit candidate lists.}
\label{fig:SNR}
\end{figure}

The transit duration, $D$, is calculated using the following relation,
\begin{equation}
D=\frac{\Rs P}{\pi a}
\label{eqn:duration}
\end{equation}
where $P$ is the orbital period, and circular orbits are assumed.
The ephemeris and the transit duration are used to determine whether the system is in
or out of transit at each time of observation, and hence $n_\mathrm{trans}$ is calculated.
Here, partially-observed transits with more than five observations are counted towards
$n_\mathrm{trans}$. The signal-to-noise ratio for each simulated planet is calculated
for each of the three different observing baselines; the results are plotted in Fig.
\ref{fig:SNR}.

The number of simulated transiting planets with $\rmsub{S}{red}$ greater than or equal to 10
was also calculated for each of the 20 fields which have at least 10 nights of observations.
Each of these fields has a different number of nights of observations, and this is reflected
in the simulation, which was conducted 100 times to reduce the problems of small number
statistics. A total of $3.72 \pm 1.60$ planets were detected from a population of $151 \pm
13$ transiting extra-solar planets. The detailed results of this simulation are shown in
Table \ref{tab:new}, and the detection rate for these fields is plotted as a function of the
number of observing nights in Fig. \ref{fig:detections}. Also plotted in Fig.
\ref{fig:detections} are the detection rates produced by the full simulation for the case of
both one and two seasons of observing. The detection rate of transiting extra-solar planets
increases linearly with the number of observing nights.

\section{Discussion}
\subsection{Red noise and the {\sc SysRem} algorithm}

The {\sc SysRem} algorithm is employed to remove four components of systematic error by
removing trends which are present in all the stars in a particular field. If the
implementation of {\sc SysRem} is changed so that the number of trends set for removal is
increased to five or more, no change in the quality of the data is observed. Despite this,
however, red noise -- as
demonstrated in \S \ref{sec:rednoise} -- is still present in the data. We postulate that
this remaining red noise does not affect stars in all parts of the field and so it cannot
be removed by {\sc SysRem}. Instead, we suggest that the surviving red noise is localised
and is most likely to be caused by variations in detector characteristics coupled with
variations in the tracking of the mount. For instance, a particular group of stars may
drift over the shadow cast on the CCD by a grain of dust on the optics, producing a
regular dimming in a small number of objects.

\subsection{Limitations of the Besan\c con-based model}

The Besan\c con-based model used here has certain limitations; it should be
noted that the simulation is based only on the observations from one of the
SuperWASP-N cameras and it is assumed that none of the stars are blended. Only
about two thirds of the generated planets have periods less than 5 days and, at
present, {\sc Hunter} only searches for planets with periods in this range. No planets,
however, with $P > 5$ days have $\rmsub{S}{red}$ greater than 10, even with 130 nights of
observations. This is because $n_\mathrm{trans}$, and therefore $\rmsub{S}{red}$, decreases with
increasing period.

\label{sec:no_planets}
The number of planets produced in the simulation is determined to an extent by random
numbers and so varies if the simulation is repeated. The simulation used here produced 329
transiting planets from a total of 355,429 stars in the magnitude range $9.5<V<13.0$, which is a fairly typical number; the simulation was run 15 times and
the number of transiting planets produced was found to be $343\pm 25$. Additionally, the
relatively small number of planets around bright stars somewhat masks the fact that the
signal to noise is a 
function of magnitude -- bright objects tend to have better signal-to-noise ratios.

\subsection{Detection efficiency}
\label{sec:obswindow}
\begin{figure}
\subfigure{\label{fig:51nights}
\includegraphics[angle=270, width=8.25 cm]{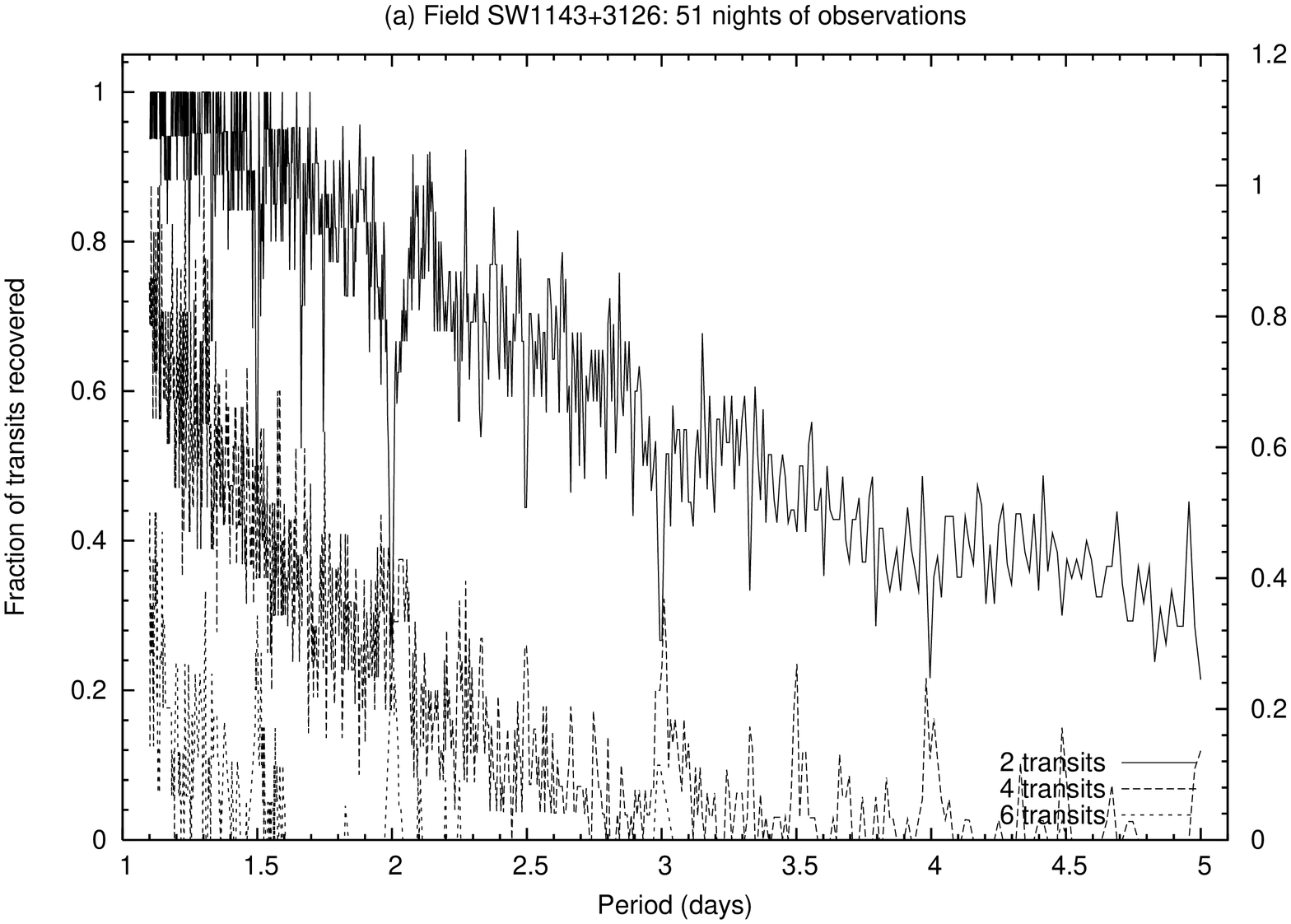}
}
\subfigure{\label{fig:80nights}
\includegraphics[angle=270, width=8.25 cm]{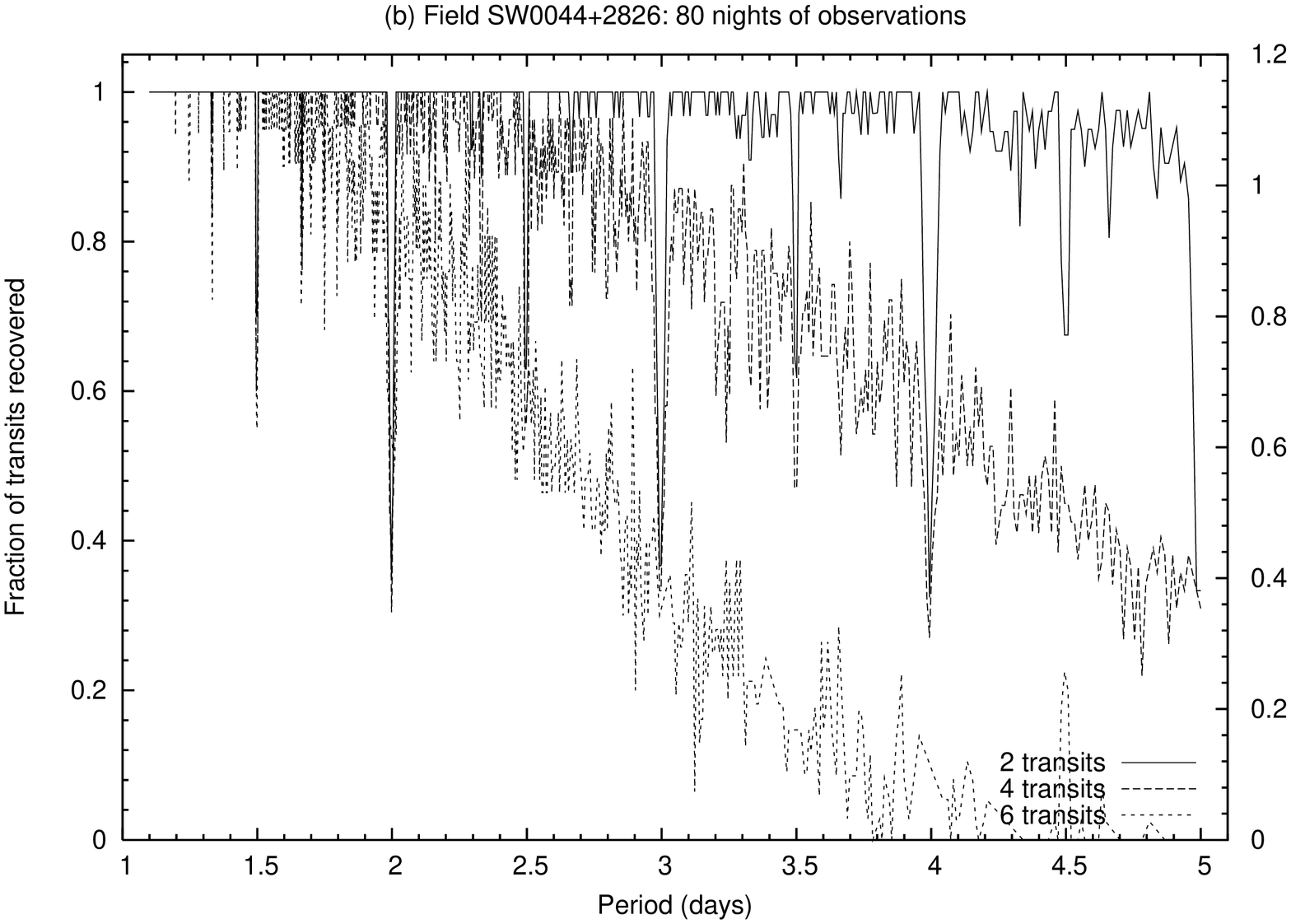}
}
\subfigure{\label{fig:130nights}
\includegraphics[angle=270, width=8.25 cm]{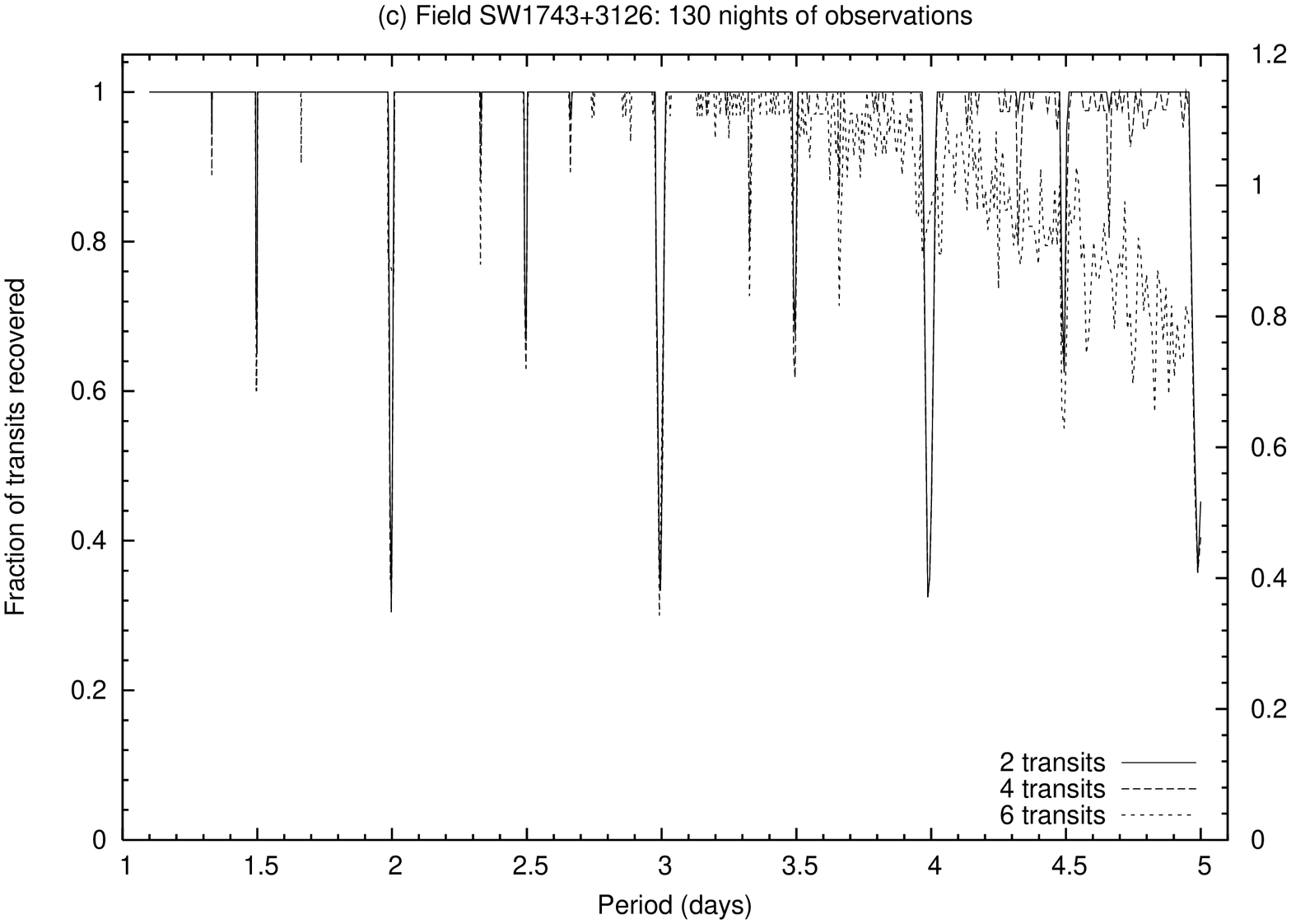}
}
\caption{Transit detection efficiency as a function of period for fields with
observations on 51, 80 and 130 nights. The solid, upper curve is for the requirement that
at least 2 transits are observed for a detection; the dashed, middle curve for 4
transits; and the dotted, lower curve for 6.}
\label{fig:transfrac}
\end{figure}

Because SuperWASP can observe only during the hours of darkness, it is impossible
to detect transits of planets with certain ephemerides, for instance a planet
that orbits with a period of exactly 1 day, where the transits occur at midday.
For a relatively modest number of observing nights, if the number of observed
transits, $n_{\mathrm{trans}}$, required to detect a planet is small (2 or 4,
say), then the detection efficiency decreases with
increasing period, but with spikes of reduced efficiency at certain
pathological periods (Fig. \ref{fig:51nights}).

Increasing the number of transits required for a detection causes the detection efficiency
to drop dramatically at most 
periods. If one requires a larger number (6 or more) of transits for
detection, then the detection efficiency is much lower, except for several pathological
periods where the detection fraction 
is such that finding planets with that period is particularly favourable.
Increasing the number of observing nights has the effect of increasing the detection
efficiency at nearly all periods (Figs.
\ref{fig:80nights},(c)).

A common property of
many of the current SuperWASP transit candidates (e.g.  \citealt{Christian-etal06}) is their large value of $n_{\mathrm{trans}}$, and
that many of the periods coincide with the narrow pathological period ranges
where there is a much greater chance of detecting a large number of transits.
The effects of red noise can be reduced by increasing the signal-to-noise of the
data by requiring that a larger number of transits $(\approx 10)$ are observed.
In order to have a reasonable chance of observing this many transits, especially
for planets with periods that are not pathologically favourable, longer
time-base observations are required. 

\begin{figure}
\includegraphics[angle=270,width=8.25cm]{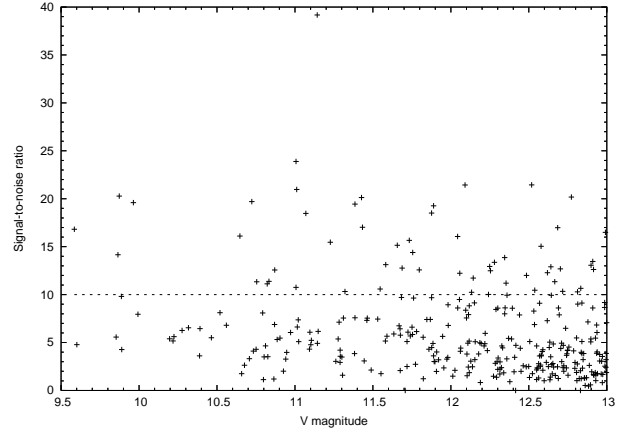}
\caption{As Fig. \ref{fig:SNR}, but for 2 seasons each consisting of 130 nights of data. 56
stars are above the $\rmsub{S}{red}$ threshold of 10.}
\label{fig:2years}
\end{figure}

The $\rmsub{S}{red}$ values of the simulated transiting planets are shown for 2 years of
observations (130 nights in 2004, and an identical 130 nights 2 years later) in Fig.
\ref{fig:2years}. This simulation shows that many more objects have a signal-to-noise
ratio above the threshold of 10 after observing for a further season.

With this in mind, the 2006 SuperWASP-N
observing season is to be dedicated to observing the same fields as in the 2004
season\footnote{SuperWASP-N did not observe during the 2005 season}. A further advantage
of this policy is that the candidate planets will have very well determined ephemerides,
which will aid follow-up work.


\begin{figure}
\includegraphics[angle=270,width=8.25cm]{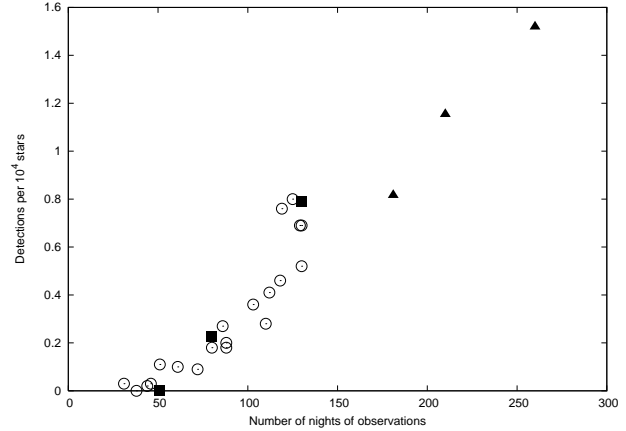}
\caption{Detection rate (a detection is defined as $\rmsub{S}{red} > 10$) of transiting
extra-solar planets versus number of nights of observations. The squares correspond to 
one season of observations, while the triangles represent an additional 130 nights of
observations in a second season. The open circles represent the 20 fields included in
Table \ref{tab:new}.}
\label{fig:detections}
\end{figure}

\begin{table*}
\centering
\caption{Simulated planetary detection rates in the twenty fields that were observed for more than ten nights by one of the SuperWASP-N
cameras in 2004.}
\label{tab:new}
\begin{tabular}{ccccccc}
\hline
Field ID & No. of nights & No. of & \multicolumn{2}{c|}{No. of stars} &
No. of planets with  & Detections  \\
 & of observations & observations & Model & Observed & $\rmsub{S}{red} \ge 10 $ in simulation & per $10^4$ stars \\
\hline
SW0043+3126 & 88 & 2962 & 6150 & 4746 &    $0.12 \pm 0.32$ & 0.20 \\
SW0044+2826 & 80 & 2851 & 5684 & 7353 &    $0.10 \pm 0.30$ & 0.18 \\
SW0143+3126 & 72 & 2251 & 6354 & 7840 &    $0.06 \pm 0.24$ & 0.01 \\
SW0243+3126 & 61 & 1646 & 7114 & 8235 &    $0.07 \pm 0.26$ & 0.10 \\
SW0343+3126 & 46 & 1286 & 9630 & 8465 &    $0.03 \pm 0.17$ & 0.03 \\
SW0443+3126 & 44 & 911 & 16062 & 8314 &    $0.04 \pm 0.20$ & 0.02 \\
SW0543+3126 & 38 & 539 & 21255 & 15021 &   $0.01 \pm 0.10$ & 0.00 \\
SW1043+3126 & 31 & 473 & 2939 & 2775 &     $0.01 \pm 0.10$ & 0.03 \\
SW1143+3126 & 51 & 813 & 2628 & 2508 &     $0.03 \pm 0.17$ & 0.11 \\
SW1243+3126 & 86 & 2157 & 2577 & 2605 &    $0.07 \pm 0.35$ & 0.27 \\
SW1342+3824 & 103 & 2578 &2482 & 2724 &    $0.09 \pm 0.32$ & 0.36 \\
SW1443+3126 & 125 & 3618 & 2988 & 3071 &   $0.24 \pm 0.51$ & 0.80 \\
SW1543+3126 & 130 & 4422 & 3899 & 3963 &   $0.27 \pm 0.47$ & 0.69 \\
SW1643+3126 & 129 & 4795 & 5388 & 6233 &   $0.37 \pm 0.59$ & 0.69 \\
SW1739+4723 & 119 & 4401 & 5361 & 8791 &   $0.41 \pm 0.58$ & 0.76 \\
SW1743+3126 & 130 & 5214 & 8712 & 11681 &  $0.45 \pm 0.65$ & 0.52 \\
SW1745+1727 & 110 & 3590 & 13700 & 17818 & $0.38 \pm 0.60$ & 0.28 \\
SW2143+3126 & 88 &  3672& 15145 & 24129 &  $0.27 \pm 0.53$ & 0.18 \\
SW2243+3126 & 118 & 4388 & 9175 & 14330 &  $0.42 \pm 0.60$ & 0.46 \\
SW2343+3126 & 112 & 3629 & 6913 & 9488 &   $0.28 \pm 0.53$ & 0.41 \\
\hline
\multicolumn{2}{l|}{Total} & & 154156 & 170090 & $3.72 \pm 1.60 $ & 0.24 \\
\hline
\end{tabular}
\end{table*}

\subsection{Comparison of detection rates with previous work}

The simulation described in this paper allows comparison with the detection rate of
transiting  HJs estimated by \cite{Brown03}, who takes no account of red noise. Of the
planets generated in the 
simulation, 63 have a transit depth deeper than one per cent, resulting in about 20
detectable planets if we assume that the fraction of transits recovered is about 0.3 for
the 38 nights of observations described by Brown. Since there are $\sim 3.55\times
10^5$ stars in our simulation, this gives a detection rate of 0.56 per $10^4$, which is
comparable to the 0.39 per $10^4$ of Brown. However, our signal-to-noise ratio analysis
suggests that, in reality, observations must be made for between 80 and 130 nights for
the detection rate of HJs to be as high as 0.39 per $10^4$ stars (Fig. \ref{fig:detections}).

\section{Conclusions}

We conclude that there is a significant component of systematic, red noise
present in data from SuperWASP-N. The {\sc SysRem} algorithm of
\cite{Tamuz-etal05}
appears highly effective at reducing the level of red noise, but fails to
eliminate it entirely. The remaining red noise is present in the data at a
level of about 3~mmag on time-scales of 2.5 hours, roughly equivalent to a
typical transit duration time. This remaining noise has a significant impact on
the efficacy of planet detection, as demonstrated by a Monte Carlo simulation based on the
Besan\c con Galaxy model. 

Our analysis reveals that if observations are conducted for only 51 nights, none of the
simulated transiting planets produces a transit detection with a signal-to-noise ratio of 10
or more. A total of $3.72 \pm 1.60$ planets which transit with $\rmsub{S}{red}~\ge~10$ are predicted
for the fields observed in 2004 by one SuperWASP-N camera, which is representative of the
other four cameras, so $18.6 \pm 8.0$ planets are predicted in total. In order to 
improve the $\rmsub{S}{red}$, and thus increase the number of detectable planets, a greater number
of transits must be observed in the data set of a particular object. This requires
observations  to be made over a longer time period.

On the basis of the transit detection rates predicted here, the SuperWASP consortium have
decided to continue observing all the fields that were monitored during 2004. We expect that
this will enhance greatly the number of planetary transit events detected at
non-pathological periods.

\section{Acknowledgements}

The WASP Consortium consists of representatives from the Universities of Cambridge (Wide
Field Astronomy Unit), Keele, Leicester, The Open University, Queens University Belfast and
St Andrews, along with the Isaac Newton Group (La Palma) and the Instituto de
Astrof\'{i}sica de
Canarias (Tenerife). The SuperWASP cameras were constructed and are operated with
funds made available from the Consortium Universities and PPARC. AMSS wishes to acknowledge the
financial support of a UK PPARC studentship.

\bibliographystyle{mn2e}
\bibliography{iau_journals,review}
\bsp
\end{document}